\documentclass[10pt,aps,prb,twocolumn,groupedaddress,longbibliography]{revtex4-1}
\usepackage{amsmath}
\usepackage{amsfonts}
\usepackage{amssymb}
\usepackage[utf8]{inputenc}
\usepackage{graphicx}
\usepackage{tabularx}
\usepackage{color}
\usepackage{caption}

\begin{document}

\title{Which model density is best in pair natural orbital local correlation theory?}

\author{R\'eka~A.~Horv\'ath}
\author{Kesha~Sorathia}
\author{Isabelle~Saint}
\author{David~P.~Tew}
\email{david.tew@chem.ox.ac.uk}
\affiliation{
University of Oxford, South Parks Road, Oxford, OX1 3QZ, UK
}

\date{\today}

\begin{abstract}
Low-scaling electron correlation theory based on the pair natural orbital approximation,
PNO-CCSD(T), has become a powerful computational tool.
Motivated by the recent discovery of large errors for organometallic molecules,
we assess the role of the model density used to discard unimportant contributions.
We find that second-order perturbation theory provides the best compromise between cost and accuracy,
but coupling between localised occupied orbitals must be accounted for. Errors in the CCSD energy are then well
below 1~kcal/mol, even for molecules with moderate multi-reference character, and the primary
remaining source of errors lies in the treatment of the (T) energy contribution.
\end{abstract}

\maketitle

\section{Introduction}

Low-scaling electron correlation theory has undergone considerable development over the past decade,
and computational approaches have emerged\cite{Riplinger:JCP144-024109,Schmitz:PCCP16-22167,Ma:WIRES8-e1371,Nagy:JCTC15-5275} 
capable of accurately computing heats of reaction and transition state energies of large molecular systems with
modest computational resources.\cite{Franzke:JCTC}
Although local approximations that exploit the short-ranged nature of electron correlation in molecules
were introduced many years ago,\cite{Sinanoglu1964,Pulay:CPL100-151,Saebo:CPL113-13,Schutz:JCP111-5691} 
in practice, the sizable errors resulting from the domain approximation\cite{Boughton:JCC14-736} 
prevented widescale application. In 2009 Neese re-introduced\cite{Neese:JCP130-114108} 
pair natural orbitial (PNO) theory,\cite{Lowdin:PR97-1474} which made it possible to use 
large local domains and eliminate instabilities arising from the domain approximation.
The PNO-CCSD(T) method (coupled cluster with singles, doubles and perturbative triples), 
has emerged as practical alternative to density functional theory for computational studies, 
particularly in organometalic chemistry, which requires accurate description of exchange, delocalisation 
and dispersion interactions.

In PNO coupled-cluster theory, each pair-correlation function $\vert \mu_{ij} \rangle $ is expanded using a
pair-specific subset of virtuals, whose number $v_{ij}$ and character are determined to recover
the correlation energy for that orbital pair to a target accuracy 
\begin{align}
\vert \mu_{ij} \rangle = \frac{1}{2} \sum_{\tilde a\tilde b}^{v_{ij}} t^{ij}_{\tilde a\tilde b} \hat E_{i\tilde a} \hat E_{j\tilde b} \vert 0 \rangle \, .
\end{align}
Here the excitation operator 
$\hat E_{i\tilde a} = \hat a^\dagger_{\tilde a\alpha} \hat a_{i\alpha} + \hat a^\dagger_{\tilde a\beta} \hat a_{i\beta}$ 
promotes an electron from a localised occupied orbital $i$ to pair natural virtual orbital
$\tilde a$. $\vert 0 \rangle$ is the Hartree--Fock reference wavefunction. 
Pair natural orbitals diagonalise the pair correlation density 
\begin{align}
&D_{ab} = 2 \sum_c [ \bar t^{ij}_{ac} t^{ij}_{bc} + \bar t^{ij}_{ca} t^{ij}_{cb} ] \\
&\sum_{ab} U^\dagger_{a\tilde a} D_{ab} U_{b\tilde b} = N_{\tilde a} \delta_{\tilde a \tilde b}
\end{align}
and provide a maximally convergent expansion for the correlation function.\cite{Lowdin:PR101-1730}
Here $\bar t^{ij}_{ac} = 2t^{ij}_{ac} - t^{ij}_{ca}$.
The size of the PNO space $v_{ij}$ for each pair 
is set by retaining PNOs with $N_{\tilde a} \ge \mathcal{T}$, where the PNO threshold $\mathcal{T}$ determines the accuracy of
the resulting correlation energy. Formally,
$t^{ij}_{ab}$ is the amplitude for the pair excitation in canonical coupled-cluster theory.
In practice, approximate amplitudes $\tilde t^{ij}_{ab}$ are used to construct a model density to determine the PNOs.\cite{MeyerJCP58-1017} The
ensuing coupled-cluster calculation is performed at low cost in the retained subset of PNOs and the contribution
to the model amplitudes from the discarded PNOs is added as an energy correction term.\cite{Neese:JCP130-114108}
The PNO approach is predicated on the assumption that the orbitals 
obtained from the approximate amplitudes are also appropriate for
representing the coupled cluster wavefunction. This letter addresses that assumption and
asks the title question: which model density is best in pair natural orbital local correlation theory?

The standard approach is to use first-order amplitudes from M{\o}ller--Plesset perturbation theory to form the model density. 
Often a semi-canonical approximation is used that neglects off-diagonal Fock matrix elements $f_i^j$ among the localised occupied orbitals.
\begin{align}
\tilde t^{ij}_{ab} = - \frac{ g^{ij}_{ab} }{ \epsilon_a + \epsilon_b - f^i_i - f^j_j }
\end{align}
where $g^{ij}_{ab} = \langle i a \vert j b \rangle$ are the electron repulsion integrals and $\epsilon_a$ the canonical 
orbital eigenvalues.
Generally, the energy from a PNO-CCSD(T) calculation $E(\mathcal{T})$ converges to the canonical result $E$ as the threshold
$\mathcal{T}$ is reduced, and the size of the PNO space increases, according to the power law\cite{Sorathia:JCP153-174112,Sorathia:JCTC}
\begin{align}
\label{eq:et}
E(\mathcal{T}) = E + A \mathcal{T}^{1/2}
\end{align}
To obtain accurate energies for the reliable prediction of chemical phenomena, 
it is necessary to ensure that the PNO truncation error $A \mathcal{T}^{1/2}$ is under control.
Neese and co-workers demonstrated\cite{Liakos:JPCA124-90} that 
mean absolute PNO truncation errors with $\mathcal{T}=10^{-7}$ are around 0.5 kcal/mol or less in PNO-CCSD(T) calculations for the
GMTKN55 data set,\cite{Goerigk:PCCP19-32184} which includes reaction energies and isomerisation energies of small
and large molecules, intermolecular and intramolecular non-covalent interaction energies and reaction barrier heights.

However, for systems with a higher proportion of static correlation, PNO truncation errors can be 
several kcal/mol, and $\mathcal{T}=10^{-8}$ or PNO extrapolation is necessary.\cite{Altun:JCTC16-6142,Sorathia:JCTC} 
This is evident from the work of Martin and co-workers\cite{Semidalas:JCTC18-883} 
on metal organic barrier heights\cite{Iron:JPCA123-3761}
and expanded porphyrin macrocycles that can interconvert between H\"uckel, figure-of-eight, and
M\"obius topologies.\cite{Sylvetsky:JCTC16-3641}

The large PNO truncation errors are an indication that the MP2 pair correlation density may fail to 
accurately model the CCSD pair correlation density for these systems,
reducing the proportion of correlation energy recovered in the selected PNO subspace 
and reducing the accuracy of the energy estimate for the discarded PNOs.

We have therefore implemented the possibility, within the \verb;pnoccsd; module in the Turbomole program package,\cite{TURBOMOLE} 
to perform a hierarchy of intermediate methods between MP2 theory and CCSD theory, 
to obtain the corresponding model density, and to perform PNO-CCSD(T) calculations using PNOs obtained from any
of these model densities. The accuracy of each model can be assessed through the size of the PNO truncation error
for computed PNO-CCSD(T) correlation energies.

If accuracy were the only criteria, then the answer to the title question would be trivial. The most
accurate model density is obviously that obtained from the canonical CCSD amplitudes, but this is of no
practical use when designing low-cost correlation methods. ``Best'' here
refers to the best compromise between accuracy and the cost incurred to obtain the model density.

The doubles amplitude equations for our chosen models are
\begin{align}
0 &= \langle \overline{\begin{subarray}{c} ab \\ ij \end{subarray}} \vert \tilde H + [ \tilde H, \hat T_2 ] + \tfrac{1}{2} [[ \tilde H, \hat T_2 ], \hat T_2 ]\vert 0 \rangle  \quad \textrm{CCSD} \\
0 &= \langle \overline{\begin{subarray}{c} ab \\ ij \end{subarray}} \vert \hat H + [ \hat H, \hat T_2 ] + \tfrac{1}{2} [[ \hat H, \hat T_2 ], \hat T_2 ]\vert 0 \rangle  \quad \textrm{CCD} \\
0 &= \langle \overline{\begin{subarray}{c} ab \\ ij \end{subarray}} \vert \tilde H + [ \tilde H, \hat T_2 ] \vert 0 \rangle  \qquad\qquad\qquad\qquad \textrm{LCCSD}\\
0 &= \langle \overline{\begin{subarray}{c} ab \\ ij \end{subarray}} \vert \hat H + [ \hat H, \hat T_2 ] \vert 0 \rangle  \qquad\qquad\qquad\qquad \textrm{LCCD}\\
0 &= \langle \overline{\begin{subarray}{c} ab \\ ij \end{subarray}} \vert \tilde H + [ \hat F, \hat T_2 ] \vert 0 \rangle  
\qquad\qquad\qquad\qquad\, \textrm{CC2} \\
0 &= \langle \overline{\begin{subarray}{c} ab \\ ij \end{subarray}} \vert  \hat H + [ \hat F, \hat T_2 ] \vert 0 \rangle  
\qquad\qquad\qquad\qquad\, \textrm{MP2}
\end{align}
where $\tilde H$ is the $T_1$ transformed Hamiltonian $e^{-\hat T_1} \hat H e^{\hat T_1}$ 
and $\langle \overline{\begin{subarray}{c} ab \\ ij \end{subarray}} \vert$
are the biorthogonal bra states for the doubles projection manifold. 
The methods CCD and LCCD are obtained from CCSD and LCCSD, respectively, by setting all $T_1$
amplitudes to zero so that $\tilde H = \hat H$. The singles amplitude equations for CC2 and LCCSD are the
same as for CCSD 
\begin{align}
0 &= \langle \begin{subarray}{c} a \\ i \end{subarray} \vert \tilde H + [ \tilde H, \hat T_2 ] \vert 0 \rangle 
\end{align}
To assess the accuracy of each model, we chose to perform calculations on the MOBH35 data set of 
barrier heights for organometallic reactions\cite{Iron:JPCA123-3761} and the POLYPYR21 data set of H\"ukel-M\"obius
porphyrin macrocycles,\cite{Sylvetsky:JCTC16-3641} which contain species with moderate amounts of static correlation and
large PNO truncation errors. Although triple-zeta quality basis sets are the minimum required to obtain reliable CCSD(T) energies for
comparison with experiment, the 
behaviour of the PNO truncation error is very similar across basis sets.\cite{Liakos:JPCA124-90,Sorathia:JCTC} 
We therefre use double-zeta quality basis sets in this work.

Two measures of accuracy are used to assess each model density. The first, $Q_d$, is a direct measure of the 
proportion of the CCSD density discarded when truncating the space using PNOs from a model density;
the second, $Q_e$, is the error in the correlation energy incurred when the selected model PNOs are used in a
PNO-CCSD(T) calculation. 

We define $Q_d(\mathcal{T})$ as the average density lost per pair using PNO truncation threshold $\mathcal{T}$
\begin{align}
Q_d(\mathcal{T}) &=  \frac{1}{n_{ij}} \sum_{i\le j} ( \mathrm{Tr} [\mathbf{D}^{ij}] - \mathrm{Tr} [\tilde{\mathbf D}^{ij}(\mathcal{T})] ) / \mathrm{Tr} [\mathbf{D}^{ij}] \\
\tilde D^{ij}_{\tilde a \tilde b} &= \sum_{ab}^{v} U^{ij}_{a\tilde a} D^{ij}_{ab} U^{ij}_{b\tilde b}
\end{align}
Here the CCSD pair correlation density $\mathbf D^{ij}$ formally has the $v\times v$ dimension of the full virtual space,
$\tilde{\mathbf D}^{ij}(T)$ is the truncated pair density that has the $v_{ij}\times v_{ij}$ dimension of the retained PNOs at
treshold $\mathcal{T}$, and 
$U^{ij}_{a\tilde a}$ is the transformation matrix from $v$ canonical virtuals to the $v_{ij}$ PNOs for pair $ij$ 
obtained from the chosen model density at threshold $\mathcal{T}$. By plotting $Q_d(\mathcal{T})$ against the average dimension $v_{ij}(\mathcal{T})$
as a function of $\mathcal{T}$ for each model density, we can compare the relative accuracy of each model.

We define $Q_e(\mathcal{T})$ as the error in the CCSD(T) correlation energy using the PNOs from the 
model density and truncation threshold $\mathcal{T}$. It is composed of three terms
\begin{align}
Q_e(\mathcal{T}) = \mathcal{E}_\text{SD}(\mathcal{T}) + \mathcal{E}_\text{T}(\mathcal{T}) - \Delta(\mathcal{T})
\end{align}
$\mathcal{E}_\text{SD}$ is the error in the CCSD correlation energy and $\mathcal{E}_\text{T}$ is the error in the 
(T) energy. The $\mathcal{E}_\text{T}$ contribution depends primarily on the truncation of the triples space through
the selection of triple natural orbitals,\cite{Riplinger:JCP139-134101} which is independent of the chosen pair density model
in the current implementation.\cite{Schmitz:JCP145-234107} Variations in $\mathcal{E}_\text{T}$ among model densities are a reflection
of the accuracy of the $T_2$ amplitudes used to compute the (T) energy. 

The $\Delta$ term corrects for $\mathcal{E}_\text{SD}$ and is
the energy estimate for the discarded PNOs based on the model amplitudes
\begin{align}
\Delta = \sum_{ij} \left( \sum_{ab}^v \tilde t^{ij}_{ab} \bar g_{ij}^{ab} 
 - \sum_{\tilde a\tilde b}^{v_{ij}} \tilde t^{ij}_{\tilde a\tilde b} \bar g_{ij}^{\tilde a\tilde b} \right)
\end{align}
Here $\bar g_{ij}^{ab} = 2g_{ij}^{ab} - g_{ij}^{ba}$. We report error statistics in units of mE$_h$ per valence electron
for total energies, and in kcal/mol for reaction energies and barrier heights.

\begin{figure}[!tbp]
   \begin{center}
   \includegraphics[width=0.95\columnwidth]{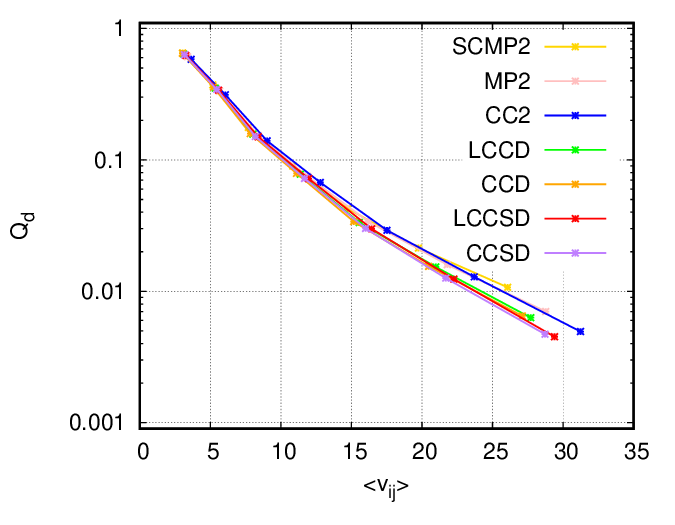}
   \end{center}
\caption{\label{fig:Qdr29} The loss of information against average number of PNOs for different model densities for
reactant 29 of the MOBH35 set}
\end{figure}

\section{Computational details}

The structures for the MOBH35 test set were taken from the supporting information of
Ref~\onlinecite{Semidalas:JCTC18-883}, where the transition state structures for reactions 11 and 12,
and all species of reaction 14 are modified from the original data set, as recommended by Dohm \emph{et al}.\cite{Dohm:JCTC16-2002}
We use the def2-SVP\cite{Weigend:PCCP7-3297} basis and density fitting basis\cite{Weigend:PCCP8-1057}
and the Stuttgart relativistic effective core potentials\cite{Dolg:CR112-403}
for molecules containing second- and third-row transition metal atoms. 
Reactant 16 converges to the incorrect state if the default extended H\"uckel orbital guess is applied and an orbital swap was
required to obtain the correct initial electronic configuration.
Canonical CCSD(T) energies for 19 of the 35 reactions are available from earlier work,\cite{Sorathia:JCTC} so we use this subset
MOBH19 for the current study. These are reactions 3, 4, 6, 7, 13, 14, 15, 16, 21, 23, 26, 27, 29, 30, 31, 32, 33, 34 and 35.

The structures for the POLYPYR21 data set were taken from the supporting information of Ref.~\onlinecite{Sylvetsky:JCTC16-3641}.
To enable direct comparison with the CCSD and CCSD(T) energies of the original work, we use the cc-pVDZ basis set\cite{Dunning:JCP90-1007} 
with the $d$ functions on the H atoms removed, and the cc-pVDZ density fitting basis.\cite{Weigend:JCP116-3175} 

All calculations were performed using the the \verb;dscf;\cite{HaeserJCC10-104} and 
\verb;pnoccsd; modules,\cite{Schmitz:PCCP16-22167,Schmitz:JCP145-234107,Tew:JCTC15-6597}
of the Turbomole package. Density fitting was not used in the Hartree--Fock calculations.
Orbitals were localised using the intrinsic bond orbital method.\cite{Knizia:JCTC9-4834} and
in all cases the multilevel approximation\cite{Masur:JCP139-164116,Schutz:JCP140-244107} were not applied.
The $Q_d$ and $Q_e$ measures were computed from near-canonical model densities obtained from the
corresponding PNO coupled cluster calculation using a very tight PNO threshold of $\mathcal{T}$=$10^{-9}$. 
Where timings are reported, these were performed on a single Intel(R) Xeon(R) Gold 6248R CPU @ 3.00GHz node with
48 cores, 380 Gb RAM and 1.8Tb SSD.

\section{Results}

\begin{figure}[!tbp]
   \begin{center}
   \includegraphics[width=0.95\columnwidth]{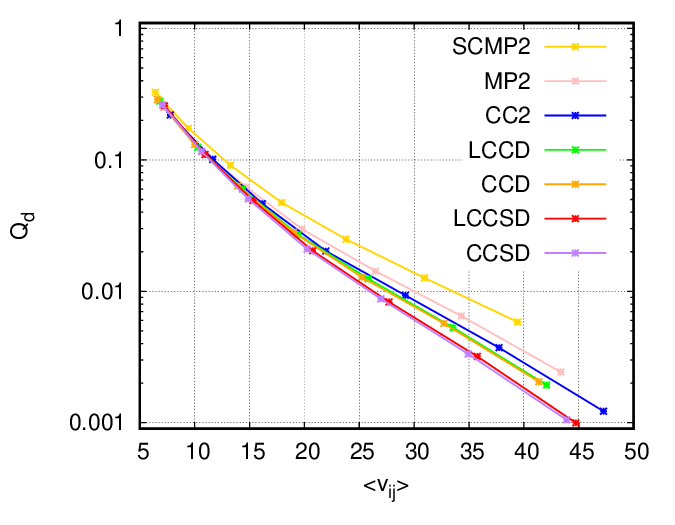}
   \end{center}
\caption{\label{fig:Qdr32} The loss of information against average number of PNOs for different model densities for
reactant 32 of the MOBH35 set}
\end{figure}

Figs.~\ref{fig:Qdr29} and \ref{fig:Qdr32} display the average information lost $Q_d(\mathcal{T})$ against the average number of PNOs $v_{ij}(\mathcal{T})$ retained with $\mathcal{T}$=$10^{-5}$--$10^{-8}$ for the hierarchy of model densities.
The two chosen molecules r29 and r32 are the extremal cases of the MOBH19 set where the SCMP2 density performs best and worst relative to CCSD, respectively. The CCSD model loses the least possible information per PNO discarded and 
is the reference against which the other methods are judged.

The LCCSD and LCCD models are only marginally inferior to CCSD and CCD, respectively.
The influence of the quadratic $T_2$ terms in the CC model are therefore small on average.
We find that they are more significant for strong pairs than weak pairs, but in most cases have less impact than the singles 
cluster amplitudes. Particulary for molecules with substantial correlation-induced orbital relaxation, 
the CCD and LCCD models lose accuracy relative to CCSD and CC2 is more accurate than MP2. 
In almost all cases MP2 is significantly better than semi-canonical MP2.
Neglecting the Fock matrix elements that couple the localised orbitals is a poor approximation.
Overall, the accuracy of the CC model densities is in line with that expected from a perturbative analysis of the 
the amplitude equations, and the magnitude of the interaction follows $f_i^j > [\hat H, T_2] > T_1 > [[\hat H, T_2], T_2]$.

\begin{figure}[!tbp]
   \begin{center}
   \includegraphics[width=0.95\columnwidth]{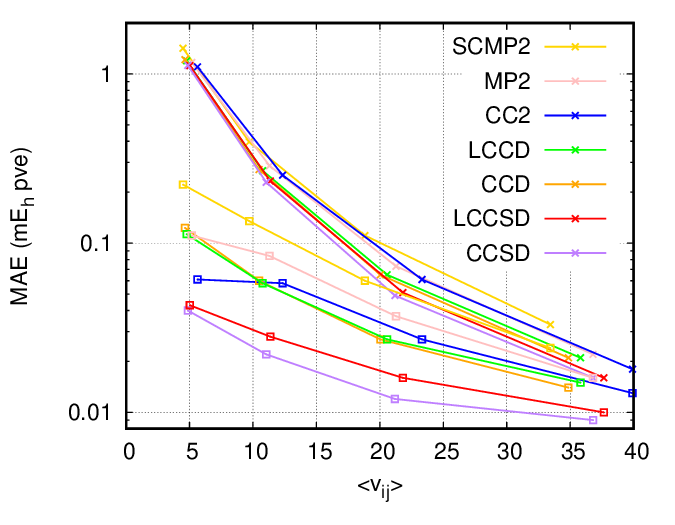}
   \end{center}
\caption{\label{fig:qesd} Mean absolute PNO truncation errors $\mathcal{E}_\text{SD}$ ({\tiny $\times$}) and $\mathcal{E}_\text{SD} - \Delta$ ({\tiny $\square$}) against the average number of PNOs for the MOBH19 set using various model densities}
\end{figure}

Fig.~\ref{fig:qesd} shows the mean absolute truncation error per valence electron in the CCSD correlation energy against
the mean number of PNOs per pair, averaged over the dat set, for $\mathcal{T}$=$10^{-5}$--$10^{-8}$. 
For each model density, we plot the mean absolute values of $\mathcal{E}_\text{SD}$ and $\mathcal{E}_\text{SD} - \Delta$, 
to ascertain the relative accuracy of the energy estimate for the discarded PNOs. The mean $\mathcal{E}_\text{SD}$ 
values for each method mirror the trend observed for $Q_d$, with CCSD$\sim$LCCSD$<$CCD$\sim$LCCD$<$CC2$\sim$MP2$<$SCMP2.

\begin{figure}[b]
   \begin{center}
   \includegraphics[width=0.95\columnwidth]{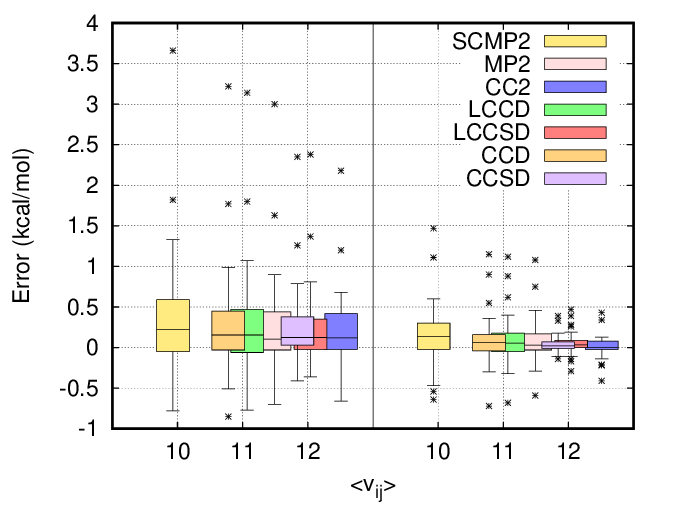}
   \end{center}
\caption{\label{fig:DEsd6} Box plots for PNO-CCSD errors $\mathcal{E}_\text{SD}$ (left) and $\mathcal{E}_\text{SD}-\Delta$ (right) 
in MOBH19 barrier heights using $\mathcal{T}$=$10^{-6}$.}
\end{figure}

A greater variation in accuracy is observed for $\mathcal{E}_\text{SD} - \Delta$, but again the same trend is observed.
However, we find that the CC2 model is significantly more accurate than MP2 when the $\Delta$ correction is taken into account.
Interestingly, for CCSD, the $\Delta$ correction does not entirely eliminate the PNO trunctation error, and
a small error remains due to relaxation of the $T_2$ amplitudes when solving the non-linear CCSD equations in the retained subspace.

\begin{table*}[!thbp]
\caption{PNO truncation error normal distributions $(\mu,\sigma_n)$ for barrier heights of the MOBH19 set in kcal/mol}
\label{tab:ev}
\begin{tabular*}{0.95\textwidth}{@{\extracolsep{\fill}} llllrrrrrrrrrrrrrrrrrrrr}
\hline\hline
     &&          && \multicolumn{2}{c}{SCMP2}  && \multicolumn{2}{c}{MP2} && \multicolumn{2}{c}{CC2} 
                 && \multicolumn{2}{c}{LCCD}   && \multicolumn{2}{c}{CCD}
                 && \multicolumn{2}{c}{LCCSD}  && \multicolumn{2}{c}{CCSD}  \\
Contr.  && $\mathcal{T}$   
&& \multicolumn{1}{c}{$\mu$}  &  \multicolumn{1}{c}{$\sigma$} && \multicolumn{1}{c}{$\mu$}  &  \multicolumn{1}{c}{$\sigma$} && \multicolumn{1}{c}{$\mu$}  &  \multicolumn{1}{c}{$\sigma$} && \multicolumn{1}{c}{$\mu$}  &  \multicolumn{1}{c}{$\sigma$} && \multicolumn{1}{c}{$\mu$}  &  \multicolumn{1}{c}{$\sigma$} && \multicolumn{1}{c}{$\mu$}  &  \multicolumn{1}{c}{$\sigma$} && \multicolumn{1}{c}{$\mu$}  &  \multicolumn{1}{c}{$\sigma$} \\
\cline{1-3} \cline{5-6} \cline{8-9} \cline{11-12} \cline{14-15} \cline{17-18} \cline{20-21}  \cline{23-24} 
$\mathcal{E}_\text{SD}$
      && $10^{-5}$ && 1.23 & 2.06 && 0.94 & 1.72 && 0.78 & 1.50 && 1.12 & 1.99 && 1.13 & 2.03 && 0.98 & 1.79 && 1.03 & 1.78 \\
      && $10^{-6}$ && 0.37 & 0.74 && 0.27 & 0.61 && 0.20 & 0.46 && 0.30 & 0.64 && 0.29 & 0.65 && 0.24 & 0.50 && 0.23 & 0.48 \\
      && $10^{-7}$ && 0.12 & 0.24 && 0.08 & 0.19 && 0.04 & 0.12 && 0.09 & 0.20 && 0.09 & 0.20 && 0.06 & 0.12 && 0.06 & 0.12 \\
      && $10^{-8}$ && 0.03 & 0.09 && 0.02 & 0.07 && 0.01 & 0.05 && 0.03 & 0.07 && 0.02 & 0.07 && 0.02 & 0.05 && 0.01 & 0.04 \\[1ex]
$\mathcal{E}_\text{SD}-\Delta$
      && $10^{-5}$ && 0.20 & 0.73 && 0.08 & 0.48 &&$-$0.03& 0.34 && 0.16 & 0.55 && 0.16 & 0.55 && 0.09 & 0.28 && 0.04 & 0.19 \\
      && $10^{-6}$ && 0.15 & 0.38 && 0.08 & 0.27 &&  0.02 & 0.14 && 0.10 & 0.30 && 0.09 & 0.30 && 0.05 & 0.14 && 0.04 & 0.10 \\
      && $10^{-7}$ && 0.08 & 0.18 && 0.03 & 0.12 &&  0.01 & 0.06 && 0.04 & 0.13 && 0.04 & 0.13 && 0.02 & 0.06 && 0.01 & 0.05 \\
      && $10^{-8}$ && 0.02 & 0.08 && 0.01 & 0.06 &&  0.01 & 0.04 && 0.02 & 0.06 && 0.01 & 0.07 && 0.01 & 0.04 && 0.01 & 0.04 \\[1ex]
$\mathcal{E}_\text{(T)}$
      && $10^{-5}$ && 0.78 & 1.41 && 0.62 & 1.09 && 0.61 & 1.05 && 0.62 & 1.14 && 0.64 & 1.17 && 0.64 & 1.09 && 0.64 & 1.09 \\
      && $10^{-6}$ && 0.41 & 0.73 && 0.28 & 0.55 && 0.28 & 0.51 && 0.29 & 0.58 && 0.30 & 0.59 && 0.29 & 0.53 && 0.29 & 0.54 \\
      && $10^{-7}$ && 0.16 & 0.31 && 0.11 & 0.24 && 0.11 & 0.22 && 0.12 & 0.25 && 0.12 & 0.26 && 0.11 & 0.22 && 0.11 & 0.22 \\
      && $10^{-8}$ && 0.05 & 0.11 && 0.04 & 0.09 && 0.03 & 0.08 && 0.04 & 0.09 && 0.04 & 0.10 && 0.04 & 0.08 && 0.04 & 0.09 \\[1ex]
$Q_e$
      && $10^{-5}$ && 0.98 & 1.84 && 0.70 & 1.34 && 0.58 & 1.06 && 0.78 & 1.52 && 0.80 & 1.56 && 0.72 & 1.21 && 0.68 & 1.14 \\
      && $10^{-6}$ && 0.56 & 1.06 && 0.36 & 0.77 && 0.30 & 0.59 && 0.38 & 0.82 && 0.39 & 0.84 && 0.34 & 0.63 && 0.33 & 0.60 \\   
      && $10^{-7}$ && 0.24 & 0.47 && 0.14 & 0.34 && 0.12 & 0.26 && 0.16 & 0.36 && 0.16 & 0.37 && 0.13 & 0.26 && 0.13 & 0.26 \\   
      && $10^{-8}$ && 0.08 & 0.19 && 0.05 & 0.14 && 0.04 & 0.11 && 0.05 & 0.15 && 0.05 & 0.15 && 0.04 & 0.11 && 0.05 & 0.11 \\[1ex]
\hline\hline
\end{tabular*}
\end{table*}

In Table~\ref{tab:ev} we report statistical measures for the distribution of PNO truncation errors in the 
barrier heights of the MOBH19 set. From the data for the CCSD model, we see that for both $\mathcal{E}_\text{SD}$
and $\mathcal{E}_\text{(T)}$ the $2\sigma$ 95\% confidence interval falls within
1 kcal/mol only for $\mathcal{T}$=$10^{-7}$ and tighter. The $\mathcal{E}_\text{SD}$ contribution is 
almost entirely removed by the $\Delta$ correction and the error in the (T) energy becomes the limiting factor.
All of the approximate models follow this same pattern. 

\begin{figure}[b]
   \begin{center}
   \includegraphics[width=0.95\columnwidth]{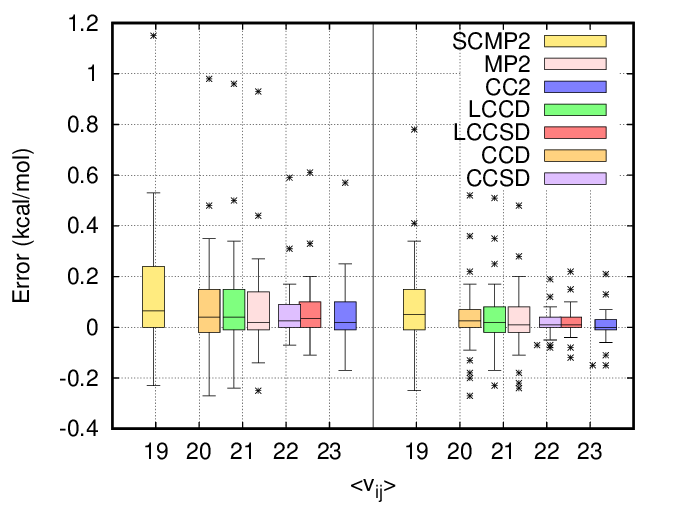}
   \end{center}
\caption{\label{fig:DEsd7} Box plots for PNO-CCSD errors $\mathcal{E}_\text{SD}$ (left) and $\mathcal{E}_\text{SD}-\Delta$ (right)
in MOBH19 barrier heights using $\mathcal{T}$=$10^{-7}$.}
\end{figure}

Fig.~\ref{fig:DEsd6} displays box plots for the PNO truncation errors in the PNO-CCSD barrier heights
of the MOBH19 set, with and without the correction for discarded PNOs, using the various model densities with
$\mathcal{T}$=$10^{-6}$. Fig.~\ref{fig:DEsd7} displays the corresponding information for $\mathcal{T}$=$10^{-7}$.
To indicate the accuracy relative to the subspace size, the data for each method is positioned along the $x$ axis 
at the average of $\langle v_{ij} \rangle$ for the transition states.

From the table and the figures, we see that for a given threshold $\mathcal{T}$, the CC2 model gives the smallest
errors, but this is because it predicts a larger density and therefore more PNOs are retained. Nevertheless, the $\Delta$
correction with the CC2 model appears to be particularly accurate. The ratio of accuracy to number of PNOs for 
barrier heights using the model densities follows the same trend as that seen for total energies and information content.
The MP2 model is remarkably accurate overall and the benefit of using higher-order CC models appears marginal, 
especially since the (T) error is the dominant contribution.

The $\mathcal{E}_\text{(T)}$ statistics are very similar for all methods. The primary factor in  $\mathcal{E}_\text{(T)}$ is the
model density used for the TNO truncation, which is independent of the PNO model density in this work. 
The error in the (T) energy due to 
errors in the $T_2$ amplitudes are smaller, but not insignificant, and give rise to some variation among the model
densities tested. We find that the maximum errors are generally smaller for CC2 and LCCSD than for LCCD and CCD.
Overall, the SCMP2 performs the worst, and the majority of the available improvement is obtained already at the MP2 level.
Using MP2 or CC2 model densities with $\mathcal{T}$=$10^{-7}$ is sufficient to reduce the maximum 
total PNO truncation errors to within 1 kcal/mol.

The cost of using a higher-order CC model to obtain an approximate density has two sources:
the overhead of performing the higher-order CC model using a tight threshold; and a cost or benefit arising
from a larger or smaller PNO set for the subsequent PNO-CCSD(T) calculation. The overhead can be reduced by
loosening the PNO threshold used to obtain the model density. 

\begin{table}[b]
\caption{Mean absulte (and maximum) PNO-CCSD errors for MOBH19 barriers in kcal/mol}
\label{tab:grid}
\begin{tabular*}{0.95\columnwidth}{@{\extracolsep{\fill}} llrrrrrr}
\hline\hline
Density                & \multicolumn{2}{c}{$\mathcal{T}=10^{-5}$} & \multicolumn{2}{c}{$\mathcal{T}=10^{-6}$} & \multicolumn{2}{c}{$\mathcal{T}=10^{-7}$} \\
\hline
SCMP2 $(10^{-9})$      &$   0.51 $&$ (2.61) $&$   0.28 $&$ (1.47) $&$  0.13 $&$ (0.78) $\\
MP2 $(0.1\mathcal{T})$ &$   0.38 $&$ (1.45) $&$   0.22 $&$ (1.20) $&$  0.09 $&$ (0.54) $\\
MP2 $(10^{-9})$        &$   0.31 $&$ (1.80) $&$   0.18 $&$ (1.08) $&$  0.08 $&$ (0.48) $\\
CC2 $(0.1\mathcal{T})$ &$   0.26 $&$ (0.78) $&$   0.14 $&$ (0.80) $&$  0.06 $&$ (0.34) $\\
CC2 $(10^{-9})$        &$   0.23 $&$ (1.11) $&$   0.08 $&$ (0.43) $&$  0.04 $&$ (0.21) $\\
\hline\hline
\end{tabular*}
\end{table}

In Table~\ref{tab:grid}
we list the mean absolute and maximum errors using the MP2 and CC2 model densities computed using a threshold of $0.1\mathcal{T}$,
compared to the near-canonical densities obtained with $10^{-9}$. The errors obtained using the SCMP2 density are included for comparison. 
Evidently, using a threshold of $0.1\mathcal{T}$ is sufficient to recover more than two-thirds of the available improvement over SCMP2
on average, but the outliers are significantly improved if the $\Delta$ correction spans the larger PNO space.

\begin{figure}[!tbp]
   \begin{center}
   \includegraphics[width=0.95\columnwidth]{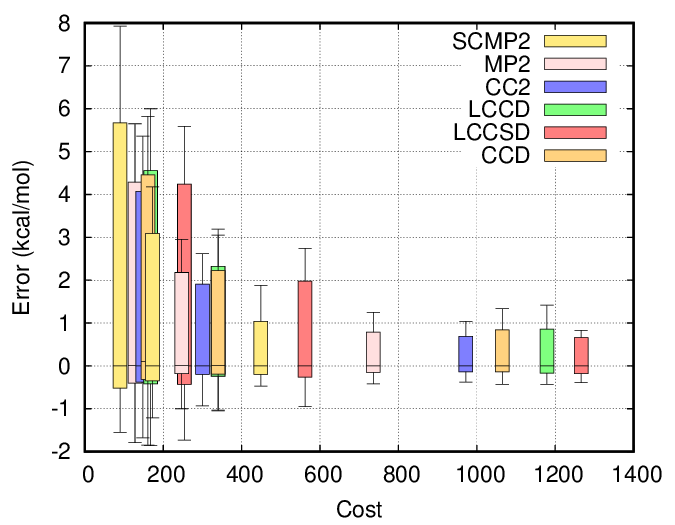}
   \end{center}
\caption{\label{fig:cost} Box plots for PNO-CCSD(T) errors of the POLYPYR21 set using different model densities
with $\mathcal{T}$=$10^{-6}$, $10^{-7}$ and $10^{-8}$.}
\end{figure}

In Fig.~\ref{fig:cost} we report the error statistics for the POLYPYR21 set using $\mathcal{T}$=$10^{-6}$, $10^{-7}$ and $10^{-8}$
and a threshold of $0.1\mathcal{T}$ for the model density. To provide a direct measure of the cost, we compute a weighted
sum of the wall times in seconds for the test set, where the weight for each molecule is the reciprocol 
of the size of the AO basis. The box plot for each model and threshold is positioned along the x axis at the value of the 
corresponding cost measure.

The PNO truncation errors for $\mathcal{T}$=$10^{-6}$ are up to 8~kcal/mol. The PNO approximation with this threshold
introduces an error so large that almost no benefit from the sophisticated coupled-cluster wavefunction model remains.
Even with $\mathcal{T}$=$10^{-7}$, errors are as much as 4~kcal/mol for the M\"obius structures.
With the SCMP2 model density and $\mathcal{T}$=$10^{-8}$, some PNO errors of 2~kcal/mol remain. However, if the MP2, CC2 or LCCSD
model densities are used, errors approach the reliability target of 1~kcal/mol. Among these, the MP2 model is the cheapest.

The timings used for the figure are only indicative, since the implementation of each approach has not been fully optimised.
Nevertheless, we note some interesting trends in the relative costs of using the model densities. For example, LCCD is more
expensive than CCD. This is because, although each iteration is cheaper, the LCCD equations take longer to converge. The
CC2 model is much more expensive than MP2, both because the number of retained PNOs is higher, and because the
singles residual contains long-ranged Fock-like terms that are expensive to evaluate. The cost of the MP2 model
can probably be reduced to close to that of the SCMP2 model by use of the Laplace transform approach.\cite{Haser:JCP96-489}

\begin{table}[t]
\caption{PNO truncation errors for CCSD energies of the POLYPYR21 set in kcal/mol}
\label{tab:poly}
\begin{tabular*}{0.95\columnwidth}{@{\extracolsep{\fill}} lrrrrr}
\hline\hline
System             & Molpro    &   Orca    &    MRCC   &    SCMP2$^a$  &    MP2$^{a,b}$    \\
\hline                                                  
24H$_\text{a}$     &$   0.1   $&$    0.0  $&$   0.0   $&$   0.0   $&$    0.0  $ \\
24H$_\text{b}$     &$   0.0   $&$    0.0  $&$   0.0   $&$   0.0   $&$    0.0  $ \\
24M                &$   0.3   $&$    0.1  $&$   0.1   $&$   0.1   $&$    0.1  $ \\
24TS$_1$           &$  -0.1   $&$    0.0  $&$   0.1   $&$   0.0   $&$    0.0  $ \\
24TS$_2$           &$   0.1   $&$    0.0  $&$   0.1   $&$   0.1   $&$    0.0  $ \\
28F                &$  -0.2   $&$   -0.6  $&$  -0.5   $&$  -0.2   $&$    0.0  $ \\
28M$_{1\text{A}}$  &$   2.0   $&$    0.5  $&$   0.3   $&$   0.3   $&$    0.3  $ \\
28M                &$   1.8   $&$    0.7  $&$   0.2   $&$   0.4   $&$    0.4  $ \\
28M$_{1\text{B}}$  &$   2.2   $&$    0.5  $&$   0.2   $&$   0.4   $&$    0.4  $ \\
28TS$_3$           &$   0.0   $&$   -0.1  $&$  -0.4   $&$   0.0   $&$    0.0  $ \\
28H                &$   0.0   $&$    0.0  $&$   0.0   $&$   0.0   $&$    0.0  $ \\
28TS$_{1\text{A}}$ &$   0.2   $&$    0.3  $&$   0.3   $&$   0.2   $&$    0.1  $ \\
28TS$_{1\text{B}}$ &$  -0.1   $&$   -0.1  $&$  -0.4   $&$   0.0   $&$    0.0  $ \\
28TS$_{2\text{A}}$ &$   2.4   $&$    1.0  $&$   0.2   $&$   0.6   $&$    0.4  $ \\
28TS$_{2\text{B}}$ &$   2.0   $&$    0.8  $&$   0.2   $&$   0.6   $&$    0.5  $ \\
32F                &$   0.0   $&$    0.0  $&$   0.0   $&$   0.0   $&$    0.0  $ \\
32H                &$  -0.7   $&$    0.5  $&$   0.4   $&$   0.3   $&$   -0.1  $ \\
32TS$_2$           &$  -0.7   $&$    0.5  $&$   0.3   $&$   0.2   $&$   -0.1  $ \\
32M$_\text{a}$     &$   1.4   $&$    0.8  $&$   1.2   $&$   0.8   $&$    0.4  $ \\
32M$_\text{b}$     &$   2.2   $&$    1.2  $&$   1.0   $&$   0.9   $&$    0.6  $ \\
32TS$_1$           &$  -0.3   $&$    0.1  $&$   0.1   $&$   0.1   $&$    0.0  $ \\[1ex]
RMS                &$   1.3   $&$    0.6  $&$   0.5   $&$   0.4   $&$    0.3  $ \\
\hline\hline
\end{tabular*} \\
$^a$ Turbomole $^b$ This work
\end{table}

In Table~\ref{tab:poly} we report the PNO truncation errors using the SCMP2 and MP2 model densities for PNO-CCSD energies
of the POLYPYR21 set, compared to those reported in Ref.~\onlinecite{Sylvetsky:JCTC16-3641} 
using the PNO-CCSD implementations in ORCA\cite{Guo:JCP148-011101} and Molpro\cite{Ma:JCTC14-198}, and using the LNO-CCSD
implementation in MRCC.\cite{Nagy:JCTC14-4193} In all cases, a threshold corresponding to $\mathcal{T}$=$10^{-7}$ is used. 
The differences among the Molpro, Orca and Turbomole PNO implementations are most likely due to the
differing domain sizes and multi-level pair approximations.
Using the full MP2 density instead of the semi-canonical density to determine the PNOs ensures that the root mean 
squared (RMS) CCSD error is 0.3~kcal/mol with a maximum error of only 0.6~kcal/mol for the challenging M\"obius structures. 

The truncation error in the PNO-CCSD(T) energies, in Fig.~\ref{fig:cost} is dominated by the (T) energy contribution.
The RMS error rises from 0.3~kcal/mol for CCSD to 1.6~kcal/mol for CCSD(T), when using the MP2 model density for the PNO
space. The maximum error rises from 0.6 to 3.0~kcal/mol. 
The truncation error can be reduced through two-point extrapolation assuming
Eq.~\ref{eq:et}. Application of two-point extrapolation of PNO-CCSD(T) energies using
$\mathcal{T}$=$10^{-6}$ and $\mathcal{T}$=$10^{-7}$ data reduces RMS and maximum errors to 0.9~kcal/mol and 1.7~kcal/mol,
respectively.
Neverthelss, the primary remaining bottleneck in accuracy for PNO calculations
on systems with moderate static correlation is treatment of the (T) energy contribution.

\section{Conclusions}

We set out to investigate the extent to which the model density used to construct the PNOs impacts the
PNO truncation errors in local correlation theory and to answer the title question 
"which model density is best?" by assessing the relative cost-benefit ratios of 
a hierarchy of model densities, starting with semi-canonical MP2 theory and ending with full CCSD.
We were motivated to address this question by the observation that large truncation errors are
observed for transition metal containing systems with moderate static correlation, where canonical coupled-cluster theory is 
expected to provide accurate predictions.\cite{Tew:JCP145-074103,Giner:JCTC14-6240,LiManni:JCTC15-1492}

We find that even when using the best possible model density, CCSD itself, the energy 
converges slowly with PNO truncation threshold for systems with a high degree of multi-reference character.
Provided that the Fock terms $f_i^j$ are not neglected in the model, there is only a modest improvement of the PNO
space when using CCSD instead of MP2 for these systems. However, the quality of the model density does greatly 
impact the accuracy of the energy correction for discarded PNOs. We find that coupling to the singles amplitudes
is more important than higher-order doubles terms in the CC model.


In the context of practical low-scaling electronic structure approaches,
the two candidates for best cost-to-accuracy ratios are therefore the MP2 and CC2 models. Due to the significant additional cost 
of computing the singles residual for CC2, our recommendation is to use the MP2 model. The application of the semi-canonical 
approximation leads to a substantial deterioration of both the PNOs and the energy correction term
and should be avoided. 

Using the MP2 density, errors in the CCSD energy are well below 1~kcal/mol with $\mathcal{T}$=$10^{-7}$
and the primary remaining error arises from the truncation of the triples space when computing the (T) energy.
Although not tested in this work, these conclusions are equally applicable to the explicitly-correlated
variants of local correlation theory.\cite{Tew:JCP135-074107,Schmitz:PCCP16-22167,InC-Tew:2021,Ma:JCTC13-4871,Pavosevic:JCP146-174108}


\bibliography{pnorefs}



\end{document}